\documentclass[12pt]{article}
\pdfoutput=1
\usepackage{geometry,enumerate,amsmath,amssymb}
\usepackage{fullpage}
\usepackage{graphicx}
\newcommand{\be}{\begin{equation}}
\newcommand{\ee}{\end{equation}}
\newcommand{\bphi}{\mbox{\boldmath $n$}}

\newcommand{\news}{\setcounter{equation}{0}\quad}
\def\ben{\begin{equation}}
\def\een{\end{equation}}
\def\beq{\begin{equation}}
\def\eeq{\end{equation}}
\def\bea{\begin{eqnarray}}
\def\eea{\end{eqnarray}}
\begin{document}
\title{
\begin{flushright}\ \vskip -2cm {\small{\em DCPT-13/17}}\end{flushright}
\vskip 2cm The dynamics of domain wall Skyrmions}
\author{
Paul Jennings and Paul Sutcliffe\\[10pt]
{\em \normalsize Department of Mathematical Sciences,
Durham University, Durham DH1 3LE, U.K.}\\[10pt]
{\normalsize Email: \quad  
paul.jennings@durham.ac.uk \quad\&\quad\  p.m.sutcliffe@durham.ac.uk}
}
\date{May 2013}
\maketitle
\begin{abstract}
It has recently been shown that Skyrmions 
with a fixed size
can exist in theories without
a Skyrme term, providing the Skyrmion is located on a domain wall.
Here
we numerically compute domain wall Skyrmions of this type, in a 
(2+1)-dimensional O(3) sigma model with a potential term. Moreover, we
investigate Skyrmion dynamics, to study both Skyrmion stability and the
scattering of multi-Skyrmions. 
We demonstrate that scattering events in which both
Skyrmions remain on the same domain wall are effectively one-dimensional, 
and at low speeds are well-approximated
by kink scattering in the integrable sine-Gordon model. 
However, more exotic fully two-dimensional scatterings are also presented, 
in which Skyrmions that are initially on different domain walls 
emerge on the same domain wall.  
\end{abstract}

\newpage
\section{Introduction}\news
Skyrmions \cite{Sk} are topological solitons in 
generalized sigma models that include a term in the Lagrangian that is
quartic in the derivatives of the field.
The role of this quartic Skyrme term is to provide a fixed finite size for 
the Skyrmion, as revealed by Derrick's theorem \cite{De}.
The original Skyrme model is a relativistic theory in (3+1)-dimensions,
where Skyrmions describe baryons within an effective field theory.
There is also a (2+1)-dimensional analogue of this theory,
known as the baby Skyrme model \cite{PSZ}. This is a generalization of the
O(3) sigma model, and has proved to be a useful 
testing ground for the study of several aspects of Skyrmions.

Non-relativistic cousins of the baby Skyrme model are also of interest in
their own right. For example, Skyrmions play a role in the
fractional quantum Hall effect \cite{SKKR} and have recently been 
experimentally observed in chiral magnets \cite{Yu}. The Skyrme term
is not appropriate for the description of these condensed matter systems
and its role is replaced by other terms that may appear in a 
non-relativistic theory. In the case of quantum Hall ferromagnets its
replacement is a non-local Coulomb interaction, and in chiral magnets 
it is the Dzyaloshinskii-Moriya interaction.
For a review of Skyrmions and their varied applications 
see \cite{book,reviewbook}.

It has recently been observed \cite{Nitta,KN} that static Skyrmions 
can exist in relativistic theories without a Skyrme term, providing the
Skyrmion is located on a domain wall. In such a theory there is no replacement
for the Skyrme term, but rather the finite fixed size of the Skyrmion
derives from the presence of the domain wall.
As theories containing higher derivative Skyrme terms can often 
be difficult to motivate physically, the presence of Skyrmions 
without the need for such a term opens up a new range of possibilities.
In particular, 
domain walls in relativistic theories are studied in both 
cosmology and high energy physics, in contexts such as  
braneworld models and supersymmetric theories, where domain walls are
important configurations that preserve a fraction of the supersymmetry.
The results of this paper may therefore find applications in these fields.
Furthermore, in the investigation of static Skyrmions, the type of 
dynamics plays no role, so applications to non-relativistic
systems in condensed matter physics are also possible.

To date, studies of Skyrmions in relativistic theories without a 
Skyrme term have been restricted to the numerical construction
of a static single Skyrmion on a domain wall in 
(2+1)-dimensions \cite{KN}.  
The form of the Skyrmion is quite different to that
familiar from theories with a Skyrme term, so it is of interest to
see what new phenomena appear. 
In this paper we present the first numerical investigations of 
Skyrmion dynamics and multi-Skyrmions in a (2+1)-dimensional theory
of this type. We provide evidence for the stability of a Skyrmion
and obtain some novel results on Skyrmion scattering.
Although our study is concerned with a planar theory,
we expect similar results to apply in (3+1)-dimensions. 

\section{The model and its static Skyrmion}\news
The theory of interest in this paper is a relativistic
(2+1)-dimensional O(3) sigma model with a potential term.
It is given by the following Lagrangian density
\be  
{\cal L}=\frac{1}{2}\partial_\mu\bphi\cdot\partial^\mu\bphi
-\frac{m^2}{2}(n_1^2+n_2^2)
+\frac{gm^2}{2}(n_1^2+n_2^2)^2n_2,
\label{lag}
\ee
where ${\bf n}=(n_1,n_2,n_3)$ is a unit vector and 
$x_\mu$ are the spacetime coordinates 
with $x_0=t,$ $x_1=x$ and $x_2=y.$
Here $m>0$ and $0<g<1$ are parameters of the theory.
This particular model has been chosen as the simplest example
of a theory that has the required properties to support a 
domain wall Skyrmion. The essential features of the theory
will become apparent below, but they are not very restrictive, so
similar results should hold in a wide range of models.

The energy corresponding to
(\ref{lag}) is   
\be
E=\int \bigg\{
\frac{1}{2}\partial_t\bphi\cdot\partial_t\bphi+
\frac{1}{2}\partial_i\bphi\cdot\partial_i\bphi
+\frac{m^2}{2}(n_1^2+n_2^2)
-\frac{gm^2}{2}(n_1^2+n_2^2)^2n_2
 \bigg\}
\,d^2x
\label{energy}
\ee
where latin indices run over only the spatial components.
The constraint $0<g<1$ ensures that the energy density is non-negative.
The potential term breaks the O(3) symmetry of the sigma model
so that only the discrete symmetries
$n_1\mapsto -n_1$ and $n_3\mapsto -n_3$ remain.
 
There are two vacuum solutions
$\bphi=(0,0,\pm 1)$ and the energy vanishes for each of these.
One of the motivations for our choice of the potential term
is that the two vacuum fields remain the same anitpodal points on the
target sphere for all values of the parameters $m>0$ and $g\in(0,1).$ 
This is not essential, and indeed in previous studies \cite{Nitta,KN}
only linear and quadratic terms appear in the potential, with the result
that the vacuum fields vary with the parameters of the theory and are 
not antipodal. The fact that our theory has fixed antipodal vacuum fields
will simplify the description of the winding structure of the Skyrmion,
so is simply for convenience.   

A field with finite energy must tend to the same vacuum solution
at spatial infinity, say ${\bf n}\rightarrow (0,0,1)$ as
$x^2+y^2\rightarrow\infty.$
An application of Derrick's theorem \cite{De} to the energy
(\ref{energy}) quickly reveals that the only finite energy
static solution is the trivial solution where ${\bf n}=(0,0,1)$ 
at all points in the plane.

The conventional approach to allow Skyrmion solutions in this
kind of theory is to introduce a Skyrme term into the
Lagrangian density (\ref{lag}), which takes the form \cite{Sk,LPZ1} 
\be
{\cal L}_{Skyrme}=-(\partial_\mu{\bf n}\times\partial_\nu{\bf n})\cdot
(\partial^\mu{\bf n}\times\partial^\nu{\bf n}).
\label{skyrmeterm}
\ee
Skyrmion solutions have been computed in such theories
with various potential terms, including the one given above with $g=0$
\cite{We}. These Skyrmion solutions are classified by the integer-valued
winding number
\be
B=-\frac{1}{4\pi}\int 
{\bf n}\cdot (\partial_x{\bf n}\times\partial_y{\bf n})\,d^2x  
\label{charge}
\ee
that counts the number of times that the field covers the target
two-sphere, ${\bf n}\cdot{\bf n}=1,$ as $(x,y)$ varies over the plane.
By definition, the topological charge $B$ is the number of Skyrmions 
in any given field configuration.
\begin{figure}
\centering
\includegraphics[width=8cm]{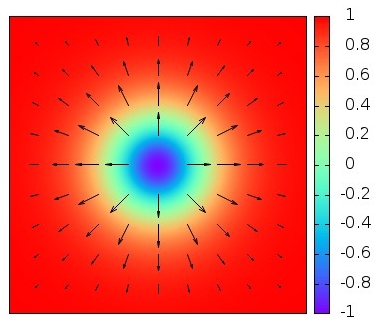}
\includegraphics[width=8cm]{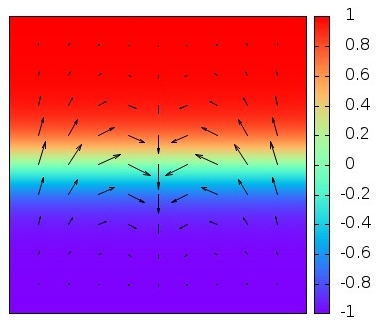}
\caption{The winding structure in the $(x,y)$ plane of a standard Skyrmion (left image)
and a Skyrmion on a domain wall (right image). The colour represents
the value of $n_3$ and the arrow indicates the length and 
direction of the two-component vector $(n_1,n_2).$} 
\label{arrows}
\end{figure}

In the presence of a Skyrme term, 
the way in which the winding is realized for the standard
single Skyrmion ($B=1$) is depicted in the left 
image in Figure~\ref{arrows}. The colour indicates the value of $n_3,$ which
varies from the vacuum value $n_3=-1$ at the centre of the 
Skyrmion (chosen as the origin of the plane) to the vacuum value 
$n_3=1$ at spatial infinity. The arrow represents the two-component
vector $(n_1,n_2),$ from which it can be seen that on the circle 
$x^2+y^2=a^2,$ with $a$ positive
and finite, this vector has a constant magnitude and takes all possible 
directions. This makes it clear that every point on the
target sphere is obtained.
The representative image shown is for the simplest case when $g=0,$ 
where the model
has a global O(2) symmetry and the Skyrmion is axially symmetric,
but the basic features of the winding structure remain the
same for $g>0,$ even though the axial symmetry is broken.  
Shortly, we shall compare this standard realization of the winding
structure with the form that it takes in the theory without a Skyrme term, 
where the Skyrmion is located on a domain wall. 

The field equations that follow from (\ref{lag}) possess a
static domain wall solution that is independent of $x$ 
and interpolates between the two vacua. Explicitly, 
consider a solution of the form
\be
\bphi=(0,\sin\theta,\cos\theta),
\label{wall}
\ee 
with $\theta(y)$ satisfying the boundary conditions that
$\theta\rightarrow \pi$ as $y\rightarrow -\infty$
and $\theta\rightarrow 0$ as $y\rightarrow \infty.$
These boundary conditions ensure that the domain wall interpolates between
the vacua since $\bphi\rightarrow (0,0,\pm1)$ as 
$y\rightarrow \pm\infty.$  
The position of the domain wall is an arbitrary parameter, which we 
choose to fix by setting $\theta(0)=\pi/2,$ so that the domain wall
lies along the $x$-axis. 

As the domain wall is independent of $x$ then its energy is infinite.
However, it has a finite tension (energy per unit length) given by 
\be
\mu=\int \bigg\{
\frac{1}{2}\theta'^2+\frac{m^2}{2}(\sin^2\theta-g\sin^5\theta)
\bigg\}\,dy.
\label{tension}
\ee
The variation of this tension yields the equation satisfied by
the static domain wall profile
\be
\theta''-{m^2}\cos\theta\sin\theta(1-\frac{5}{2}g\sin^3\theta)=0.
\label{wallprofile}
\ee
\begin{figure}
\centering
\includegraphics[width=8cm]{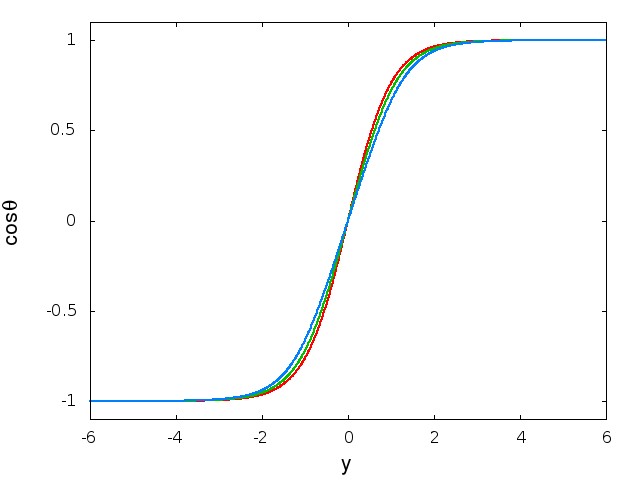}
\includegraphics[width=8cm]{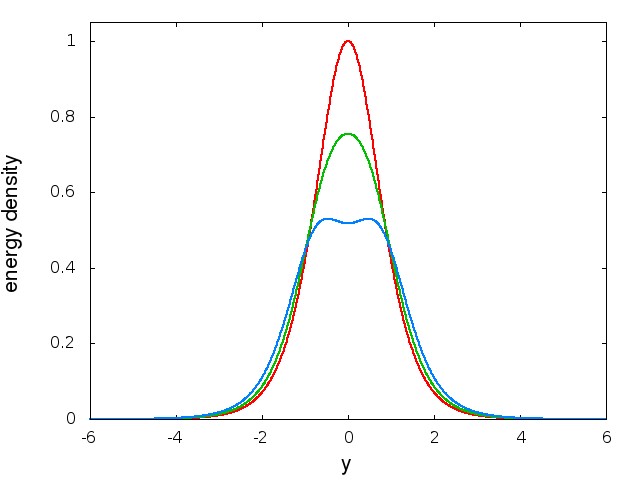}
\caption{The left image is the field $\cos\theta$ of the wall 
and the right image is the energy density. Plots for three
values of the parameter $g$ are shown: $g=0$ (red curves), 
$g=0.25$ (green curves); $g=0.5$ (blue curves). 
}
\label{fig-wall}
\end{figure}
If $g=0$ then the domain wall solution has the simple 
expression 
$n_3=\cos\theta=\tanh(my),$ with tension $\mu=2m.$
For $g>0$ the domain wall equation is hyperelliptic and does not have an 
elementary explicit solution.
However, it is a simple matter to obtain the domain wall solution numerically.
In Figure~\ref{fig-wall} we plot the domain wall field $\cos\theta$ 
(left image) and the energy density (the integrand in (\ref{tension}))
for three different values of the parameter $g.$
We see that the field does not appear to be very sensitive to the value
of $g$ although there is a more significant change in the corresponding
energy density.
 
Given a domain wall we may
dress this configuration by imposing a winding of the 
$(n_1,n_2)$ components along the domain wall. As we shall see,
this produces a domain wall Skyrmion.
Ignoring the back-reaction on the domain wall, we can use the
domain wall field $\theta(y)$ to 
obtain an effective (1+1)-dimensional theory for the Skyrmion
dynamics along the wall. 
Explicitly, we assume $\varphi(x,t)$ and consider a field of
the form
\be
\bphi=(\sin\varphi\,\sin\theta,\cos\varphi\,\sin\theta,\cos\theta).
\label{1skyrmion}
\ee
If we restrict to the domain wall ($y=0$), where $\theta=\pi/2,$ and 
substitute (\ref{1skyrmion}) into the Lagrangian density (\ref{lag}) 
the result is 
\be  
{\cal L}_{SG}=\frac{1}{2}(\partial_t\varphi)^2-\frac{1}{2}(\partial_x\varphi)^2
+\frac{gm^2}{2}\cos\varphi
-\frac{m^2}{2},
\label{lagsg}
\ee
which
is the Lagrangian density of the (1+1)-dimensional integrable
sine-Gordon model (up to the addition of an irrelevant constant).
The vacua of this theory are given by $\varphi$ an
integer multiple of $2\pi,$ corresponding to the fact that along the
domain wall ($y=0$) where $n_3=0,$ 
we have that ${\bf n}\rightarrow(0,1,0)$ as $x\rightarrow\pm\infty.$
Given vacuum boundary conditions,
$\varphi\rightarrow 2\pi N_\pm$ as $x\rightarrow \pm\infty,$
there is a conserved kink number
$N=N_+-N_-.$
The static sine-Gordon kink with $N=1$ is given by
\be
\varphi=4\tan^{-1}e^{m\sqrt{\frac{g}{2}}x},
\label{1kink}
\ee
where we have positioned the kink at the origin.

Rather than restricting the Lagrangian density to the domain wall, 
an alternative method to derive an effective (1+1)-dimensional theory
is to integrate over the spatial dimension perpendicular to the wall.
This also produces a sine-Gordon theory, though with a slightly
different scale, with the parameters in the sine-Gordon Lagrangian given
by integrals involving the wall profile function
$\theta(y).$ This approach is not as tractable as simply restricting to
the wall, because of the lack of a simple explicit expression for
$\theta(y).$ However, we have numerically obtained these parameters and
compared the two effective theories. They are similar, but it turns out
that the first approach yields a slightly better approximation to the 
full theory, as described later.     

Using the kink function (\ref{1kink}) 
in the field (\ref{1skyrmion})
gives a planar field configuration that describes a kink on a domain wall. 
Although this is not a static solution, we shall see 
that there is a static solution that is close to this field configuration.
It is easy to verify that this field has Skyrmion number $B=1,$ 
where $B$ is given by the integral expression (\ref{charge}).

The way in which the unit winding around the sphere is realized is 
depicted in the right image in Figure~\ref{arrows}. 
Along lines parallel to the $y$-axis, $n_3$ increases 
monotonically from $-1$ to $+1,$ and the vector $(n_1,n_2)$ has a
constant direction. Along lines parallel to the $x$-axis, 
$n_3$ is constant and hence the vector $(n_1,n_2)$ has constant magnitude,
but rotates through one revolution, being asymptotically vertical at
both ends of the line. It is again clear that every point on the
target sphere is realized, although the structure of the winding is quite
different to that of the standard Skyrmion. The location of the Skyrmion
has the natural definition as the point in space at which the field takes 
the value $(0,-1,0).$ This agrees with the position of the kink
on the domain wall.

\begin{figure}
\centering
\hbox{\includegraphics[width=8cm]{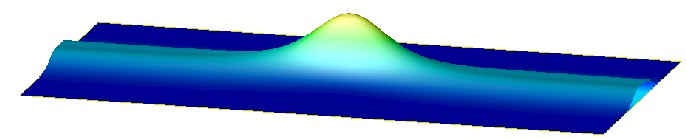}
\includegraphics[width=8cm]{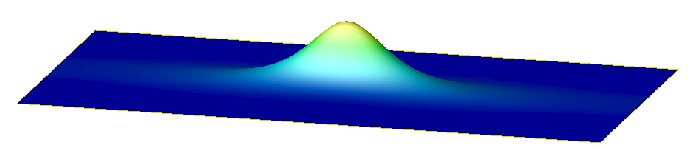}}
\caption{
The energy density (left image) and the topological charge
density (right image) of the static Skyrmion on a domain
wall.
}
\label{ebden}
\end{figure}

To obtain the true static Skyrmion we take the 
field (\ref{1skyrmion}) as an initial condition for a gradient
flow energy minimization algorithm using the static version
of the energy (\ref{energy}).
The mixed boundary condition has the Dirichlet component 
${\bf n}\rightarrow(0,0,\pm 1)$ as $y\rightarrow\pm\infty,$ 
and the Neumann component $\partial_x\bphi\rightarrow 0$ as
$x\rightarrow\pm\infty.$ 
The constant $m$ in the theory (\ref{lag}) simply determines an 
overall scale and may be set to unity
without loss of generality. For the numerical simulations presented
in this paper we fix the generic parameter value $g=\frac{1}{2}.$ 
Computations for other values of this parameter have also been performed,
and the qualitative features remain unchanged. 
Our numerical grids have a lattice spacing $\Delta x=\Delta y=0.08$ 
and the smallest grids used contain $300\times 150$ lattice points 
in the $(x,y)$ plane, with larger grids used for simulations
involving multi-Skyrmions.
% with a corresponding computational domain $[-12,12]\times[-6,6].$
Spatial derivatives are approximated using 
a fourth-order accurate finite difference scheme. 
The Dirichlet and Neumann boundary conditions, 
defined above at spatial infinity,
are applied at the boundary of the numerical grid.

In Figure~\ref{ebden} we present the result of a numerical
computation of the static Skyrmion by plotting the energy density
(the integrand in (\ref{energy})) in the left image and
the topological charge density (the integrand in (\ref{charge}))
in the right image. 
The domain wall is not visible in the plot 
of the topological charge density because this vanishes for a 
domain wall. This plot therefore highlights only the Skyrmion.

The domain wall is clearly evident in the
energy density plot, with the Skyrmion located on the wall and producing
a peak in the energy density. 
As $g\rightarrow 0$ the width of the domain wall tends to 
a non-zero value proportional to $1/m.$ However, for small $g$
the width of the kink is much larger than this, 
as it scales like $1/(m\sqrt{g}).$ 
The Skyrmion therefore appears as a structure that is
stretched in the direction of the domain wall, although this stretching 
effect is reduced as $g$ increases towards 1.

\begin{figure}
\centering
\includegraphics[width=8cm]{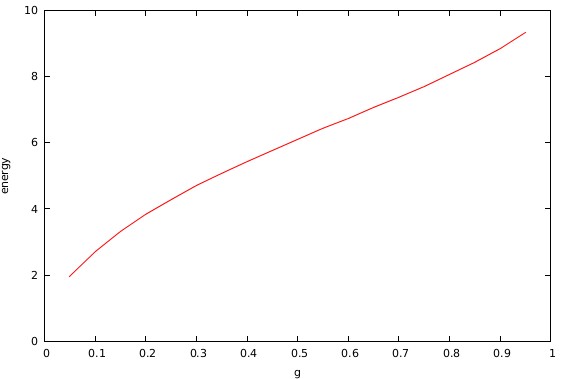}
\caption{
The energy of the static Skyrmion as a function of the parameter $g.$
}
\label{evsg}
\end{figure}

The energy of a domain wall is infinite but the additional contribution
due to the presence of a Skyrmion is finite. We may therefore define
the energy of a Skyrmion as the difference between the energies of
the domain wall Skyrmion and the pure domain wall solution.
In Figure~\ref{evsg} we plot the static Skyrmion energy as a function of the
parameter $g.$ This energy increases with $g$ and is 
approximately linear for most of the range of $g.$

\section{Skyrmion dynamics}\news
To investigate Skyrmion dynamics we numerically solve the 
(2+1)-dimensional nonlinear
wave equation that follows from the variation of the Lagrangian
density (\ref{lag}). The spatial aspects of the algorithm are as
described in the previous section on static Skyrmions, and the
time evolution is simulated using a 
fourth-order Runge-Kutta method with a timestep $\Delta t=0.02.$

\begin{figure}
\centering
\includegraphics[width=12cm]{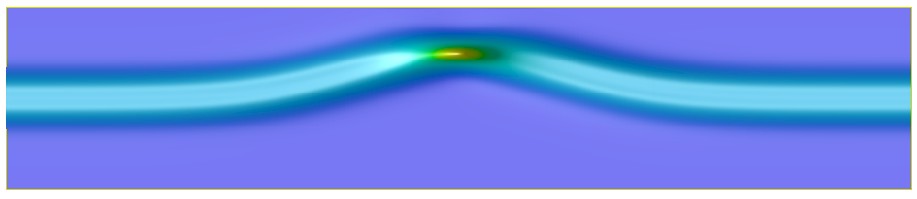}\includegraphics[width=0.85cm]{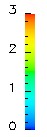}
\includegraphics[width=12cm]{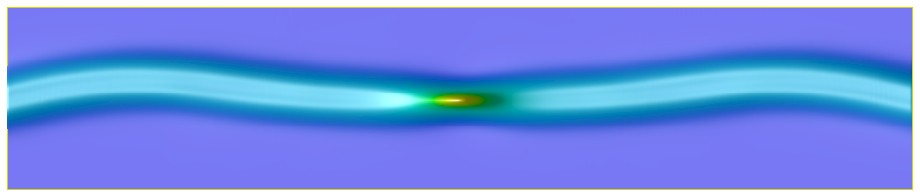}\includegraphics[width=0.85cm]{pdwlegend}
\includegraphics[width=12cm]{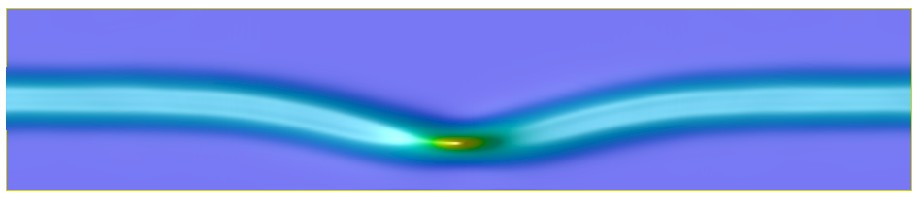}\includegraphics[width=0.85cm]{pdwlegend}
\includegraphics[width=12cm]{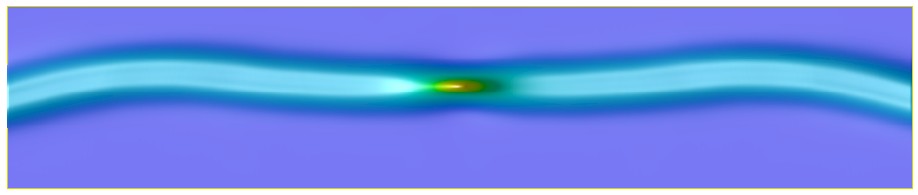}\includegraphics[width=0.85cm]{pdwlegend}
\caption{Energy density plots at increasing times 
(from top to bottom $t=0,25,60,100$)
for a perturbed domain wall Skyrmion.}
\label{pdw}
\end{figure}

To test the stability of the Skyrmion we have examined the evolution
of a number of initial conditions in which a perturbation is applied
to the static solution. A typical
example is presented in Figure~\ref{pdw}, in which a wiggle is 
introduced into the domain wall. The initial wiggle results in an
oscillation of the domain wall, but the Skyrmion remains intact on
the wall during this motion. The amplitude of the oscillation slowly
decreases as the wall radiates. This radiation can escape from the system
because of the Neumann boundary condition, and eventually the static
domain wall Skyrmion is recovered.
This, and similar results, provide strong evidence for the stability
of the domain wall Skyrmion. 

\begin{figure}
\centering
\includegraphics[width=12cm]{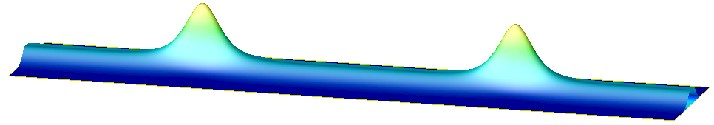}
\includegraphics[width=12cm]{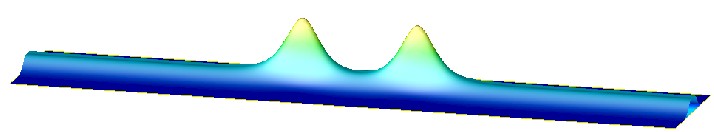}
\includegraphics[width=12cm]{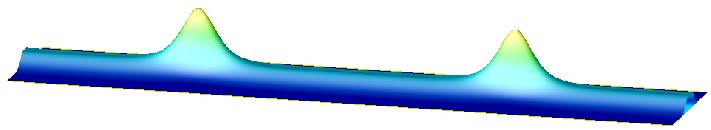}
\caption{Energy density plots at increasing times 
(from top to bottom $t=0,54,108$)
for the scattering of two Skyrmions. Each Skyrmion has an initial
velocity $v=0.2$ towards the other Skyrmion. }
\label{scat}
\end{figure}

Viewing the domain wall Skyrmion as a sine-Gordon kink on
a domain wall suggests that two Skyrmions are repulsive, since
there is a repulsive force between two sine-Gordon kinks \cite{PSk}.
This is indeed the case, as verified by evolving an initial
condition that contains two well-separated Skyrmions on the domain
wall. Furthermore, a comparison between the dynamics of two domain wall
Skyrmions and two sine-Gordon kinks reveals a good agreement for 
scattering at low speeds. 
An example scattering event is shown in Figure~\ref{scat},
in which each Skyrmion is Lorentz boosted with an 
initial velocity $v=0.2$ towards the
other Skyrmion. 
The Skyrmions approach to a minimal separation and bounce back, due to
the repulsive force, in an almost elastic collision that generates
very little radiation. 

\begin{figure}
\centering
\includegraphics[width=5cm]{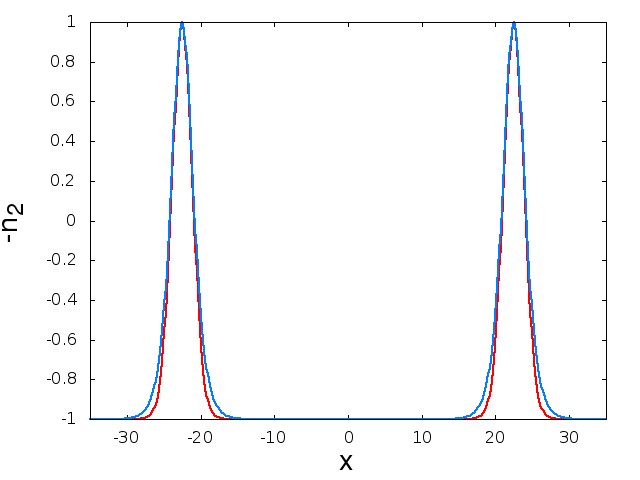}
\includegraphics[width=5cm]{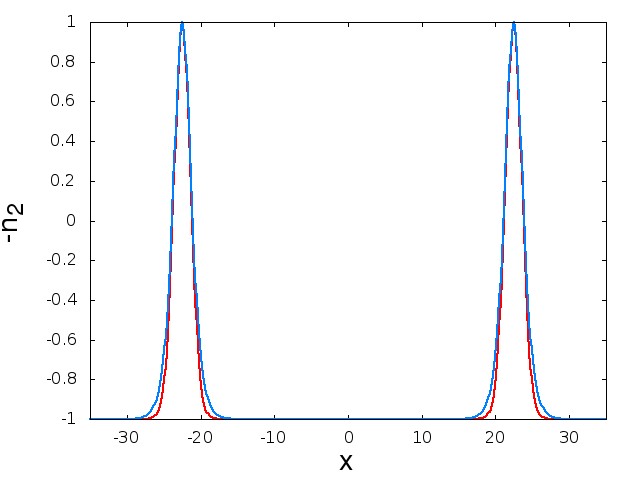}

\includegraphics[width=5cm]{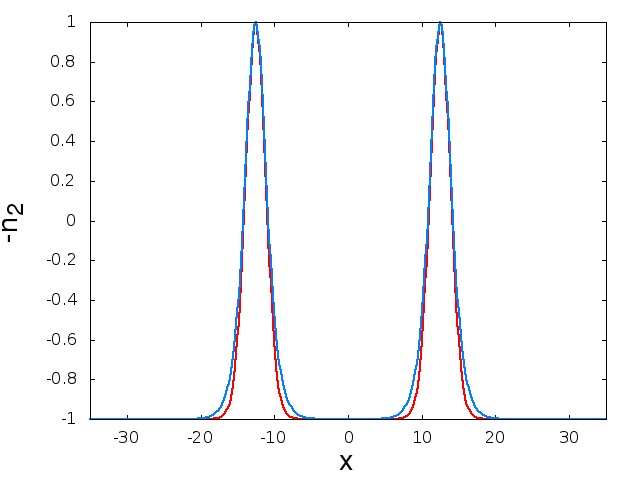}
\includegraphics[width=5cm]{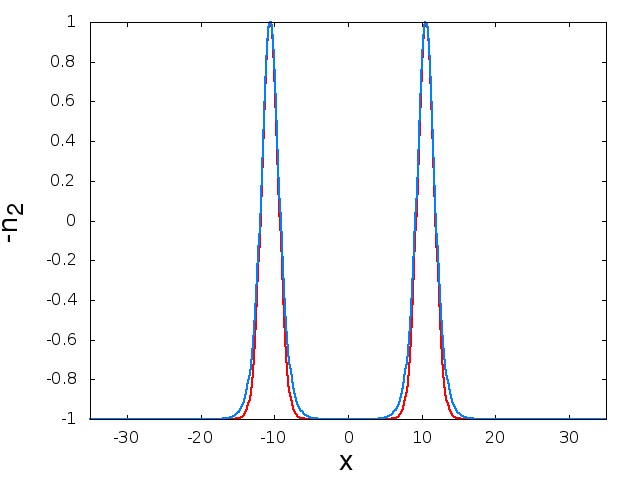}

\includegraphics[width=5cm]{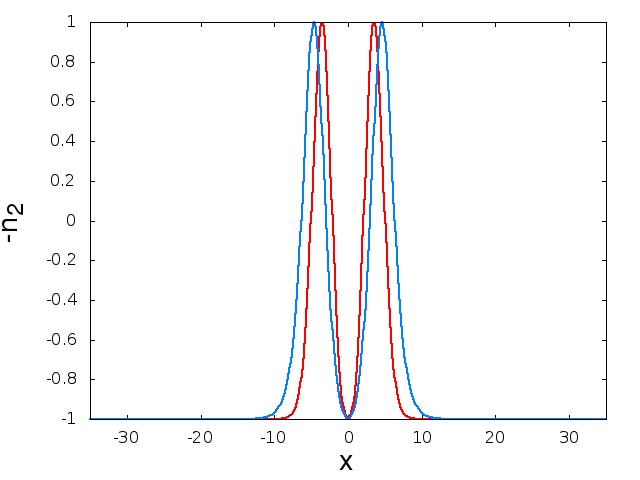}
\includegraphics[width=5cm]{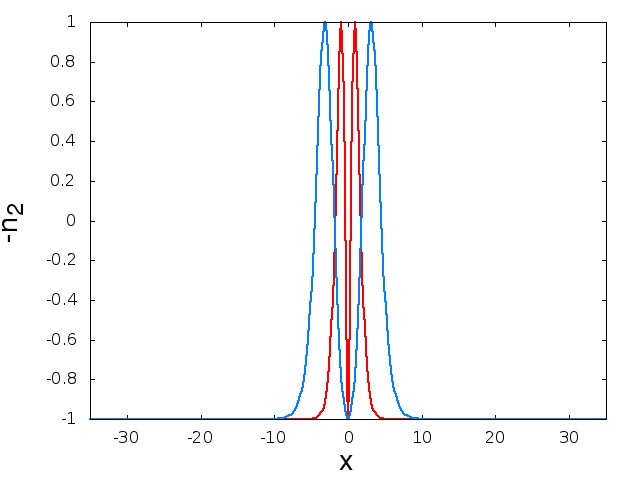}

\includegraphics[width=5cm]{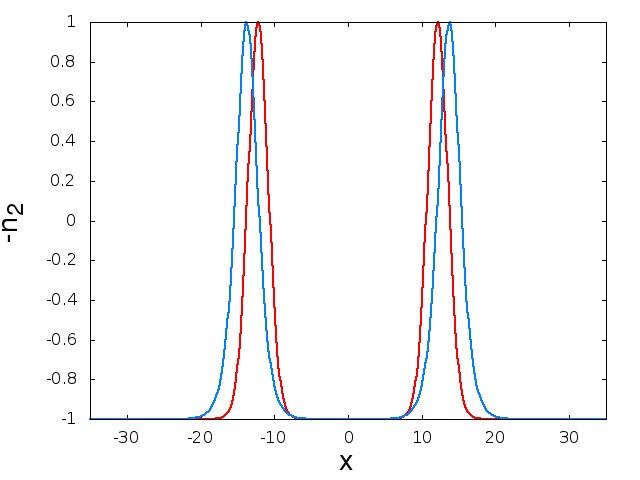}
\includegraphics[width=5cm]{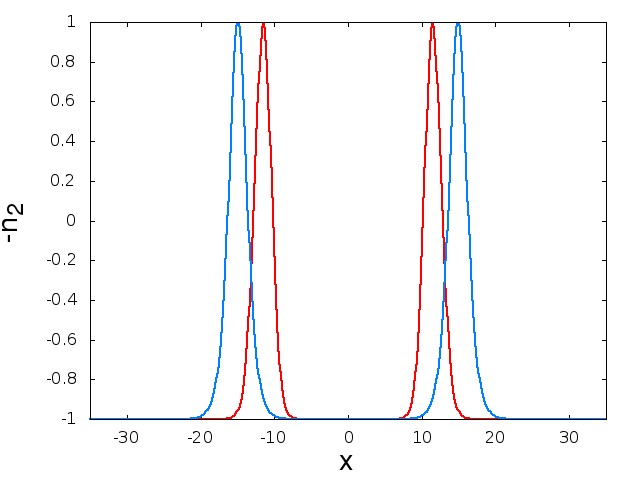}

\includegraphics[width=5cm]{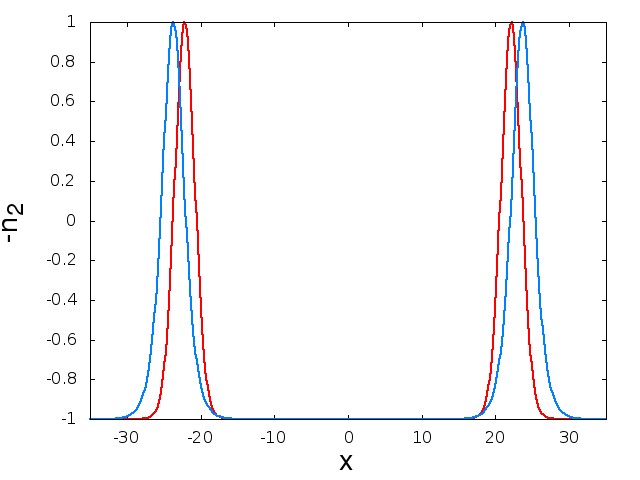}
\includegraphics[width=5cm]{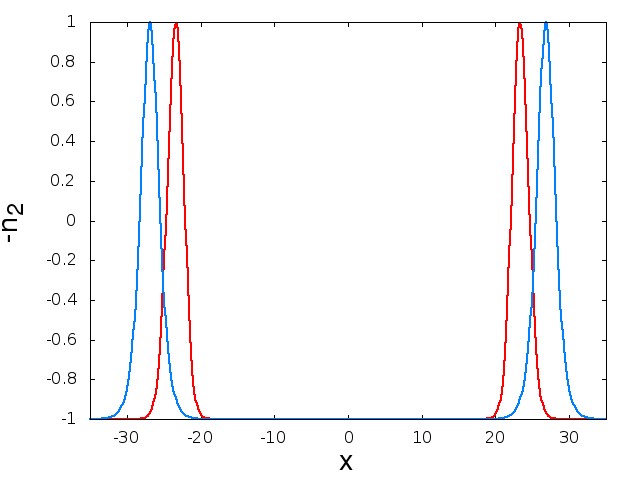}

\caption{The field $-n_2$ (red curves) along the domain wall $y=0$ at 
increasing times (from top to bottom) for an initial configuration in
which each Skyrmion has an initial velocity $v$ towards the other Skyrmion.
The associated sine-Gordon approximation is also shown (blue curves).
The images on the left are for $v=0.2$ and times $t=0,50,100,150,200.$
The images on the right are for $v=0.6$ and times $t=0,20,40,60,80.$
}
\label{slowfast}
\end{figure}

We now compare this Skyrmion scattering with kink scattering in the
integrable sine-Gordon theory (\ref{lagsg}).
In Figure~\ref{slowfast} the red curves in the images on the left
show $-n_2$ along the line $y=0$ (ie. along the 
domain wall) at increasing times during this scattering.
The blue curves display $-\cos\varphi$ 
for the corresponding exact sine-Gordon 2-kink scattering
solution
\be
\varphi=4\tan^{-1}\bigg(
\frac{v\sinh(\gamma m\sqrt{\frac{g}{2}} x)}
{\cosh(\gamma  m\sqrt{\frac{g}{2}}v(t-t_0))}\bigg),
\ee
where $\gamma=1/\sqrt{1-v^2}.$ Here $t_0$ is a constant that determines
the kink separation at time $t=0,$ and is fixed to match with the
initial conditions of the field theory simulation.

The images on the left in Figure~\ref{slowfast} reveal a good 
agreement between the Skyrmions in the field theory simulations and the 
kinks in the effective theory.
The main difference is that the minimal separation between the Skyrmions
is slightly less than the minimal separation between the kinks.
The Skyrmions get a little closer together than the kinks because as they
squeeze together they begin to explore the extra dimension.
This results in a slight increase in the time delay before the Skyrmions
bounce back. This is evident in the images at later times, where the
red curves are slightly inside the blue curves.
This effect becomes more pronounced as the initial speed $v$ increases,
as this provides more energy to explore the extra dimension.
The images on the right in Figure~\ref{slowfast} show a similar
comparison for a scattering at the increased speed $v=0.6.$ 
The closer approach of
the Skyrmions is now more apparent, as is the increased time delay. 

\begin{figure}
\centering
\hbox{\includegraphics[width=5.4cm]{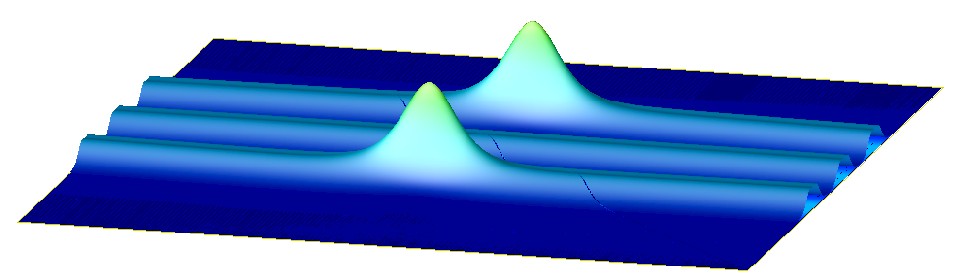}
\includegraphics[width=5.4cm]{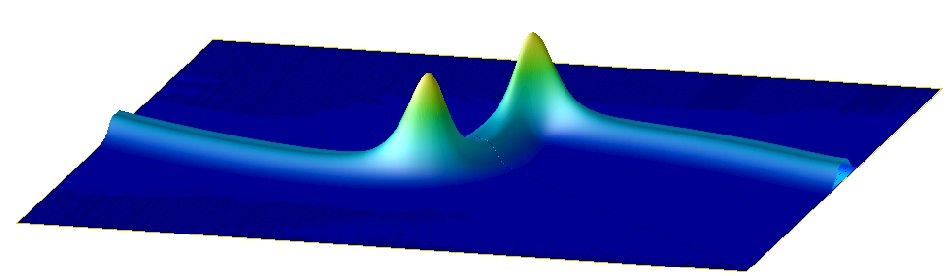}
\includegraphics[width=5.4cm]{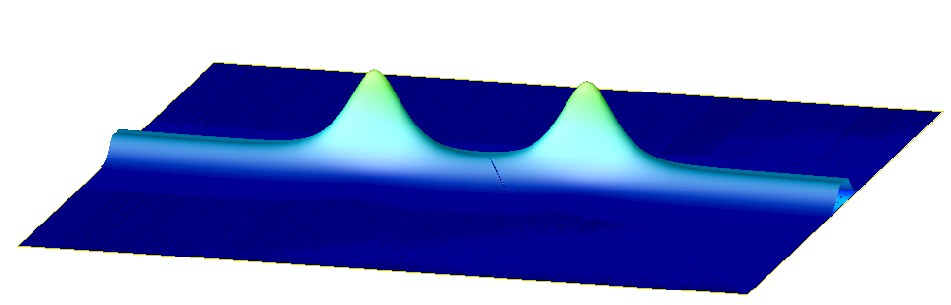}}
\caption{
Energy density plots at increasing times for the evolution 
(with damping) of two Skyrmions that are initially on different 
domain walls.
In the left image ($t=0$) the two Skyrmions have initial
positions $(-1,-7)$ and $(1,7).$ 
There are  walls along $y=\pm 7$ and an anti-wall along $y=0.$
All time derivatives are initially zero.
In the middle image ($t=60$) the Skyrmions have survived the 
wall anti-wall annihilation process. In the right image
($t=135$) the Skyrmions are moving apart along the remaining
straight wall.  
}
\label{3dw}
\end{figure}

The analogue of other phenomena that appear in the sine-Gordon theory, 
for example breathers, could also be investigated for
domain wall Skyrmions. However, we now turn to Skyrmion scattering events 
that are beyond the realm of the effective (1+1)-dimensional theory, by
considering a fully planar process in which two Skyrmions are initially 
on different domain walls. 

The vacuum structure of the theory 
is not compatible with a field configuration in which two domain
walls are placed next to each other. However, a field configuration
in which walls and anti-walls alternate is allowed. An anti-wall
is obtained from a wall by the replacement $n_3\mapsto -n_3.$  
The simplest arrangement that allows Skyrmions to be placed on different
walls involves an initial setup in which an anti-wall is sandwiched 
between two walls that each contain a Skyrmion.   

An example of such an initial configuration is displayed in the left
image in Figure~\ref{3dw}, in which there are walls along
$y=\pm 7$ and an anti-wall along $y=0.$ There is a Skyrmion on each
wall, with initial Skyrmion positions given by $(-1,-7)$ and $(1,7).$
All initial time derivatives are set to zero, so the kinetic energy
vanishes at the start of the simulation. 
It is important that there is an excess of walls over anti-walls, since
a wall and an anti-wall will generally annihilate.
Obviously, a wall is required in the final state if domain wall Skyrmions 
are to be present. As the wall anti-wall annihilation process will 
generate a large amount of kinetic energy, a damping term is included in
the dynamics to dissipate some of this kinetic energy. This aids the
visualization of the Skyrmions (as otherwise this contribution to the 
energy density can swamp that of the Skyrmions) and also reduces any
numerical difficulties associated with this radiation reflecting from the
boundary of the grid. 

The attraction between a wall and an anti-wall generates a motion in
which the two walls move towards the anti-wall. However, the fact
that the Skyrmion on the wall along $y=7$ has a positive $x$ coordinate
results in a slightly reduced speed of the wall at positive values of $x$
compared to the equivalent negative values.
A similar effect takes place for the other wall, but as the 
offset is in the reverse direction then the opposite side of the wall
is the portion with a slightly reduced speed. 
The net result is that for positive $x$
the anti-wall annihilates with the wall which initially has a
 negative value of $y,$  whereas for negative $x$ the anti-wall annihilates 
with the wall with an initial positive value of $y.$ 
The outcome is that two half-walls survive
that are joined by Skyrmions at their ends 
(see the middle image in Figure~\ref{3dw}). 
The Skyrmions survive the process in which the
two half-walls straighten into a single wall, and subsequently the
Skyrmions move apart along the remaining straight wall, due to 
their repulsive interation
(see the right image in Figure~\ref{3dw}). 
The fact that the Skyrmions survive this complicated and quite violent
process is further evidence in support of their stability.

We have performed similar simulations with other choices of initial
positions and we have also varied the strength of the damping term,
but the qualitative features remain unchanged. If no initial offset
is introduced into the $x$ components of the Skyrmion positions then
there is nothing to break the symmetry of the configuration, and hence
the Skyrmions no longer separate along the $x$-axis. 
However, this highly symmetric situation is unstable and even tiny 
numerical effects can be enough to break the symmetry and yield
the same kind of scattering as just described.

\section{Conclusion}\news
In this paper we have studied planar Skyrmions in a relativistic theory without
a Skyrme term. In particular, we have presented the first results on
the dynamics of such domain wall Skyrmions. We have investigated both
Skyrmion stability and multi-Skyrmion scattering. The latter have
been compared to kink scattering in the sine-Gordon model, which is an
effective (1+1)-dimensional theory applicable to evolution in which
all Skyrmions remain on the same wall. More exotic phenomena have also
been presented, in which Skyrmions on different domain walls can emerge
on the same wall. These examples demonstrate that the dynamics of domain
wall Skyrmions can be very complex and quite different from the usual
dynamics found for Skyrmions in theories with a Skyrme term.

For numerical simplicity, we have considered a planar theory in this
work, but we expect similar results to apply in related (3+1)-dimensional
models. Examples of suitable (3+1)-dimensional theories are discussed
in \cite{Nitta}. It might be of interest to extend the kind of numerical work
described in this paper to these higher-diemsnional theories.

Obtaining physical theories that include a Skyrme term is often
problematic, as higher-derivative terms tend not to arise naturally.
In contrast, domain walls are ubiquitous in many areas of cosmology,
 high-energy physics and condensed matter physics.
It is therefore of considerable interest to determine the properties
of domain wall Skyrmions as there are a wide range of opportunities 
for potential applications. Some of these are currently under investigation.

\section*{Acknowledgements}
PJ is supported by an STFC studentship.
PMS acknowledges funding from EPSRC under grant EP/K003453/1 and 
STFC under grant ST/J000426/1.

\end{document}